\documentclass[12pt]{article}
\usepackage{epsfig}
\usepackage{color}
\usepackage{lineno}

\begin{document}

\begin{center}

{\Large \bf{Search for alpha decay of naturally occurring osmium
nuclides accompanied by gamma quanta}}

\vskip 1.0cm

{\bf P.~Belli$^{a,b}$,
 R.~Bernabei$^{a,b,}$\footnote{Corresponding
author at: Dipartimento di Fisica, Universit\`{a} di Roma ``Tor
Vergata'', I-00133 Rome, Italy. E-mail address:
rita.bernabei@roma2.infn.it (R.~Bernabei)},
 F.~Cappella$^{c,d}$,
 V.~Caracciolo$^{a,b,e}$,
 R.~Cerulli$^{a,b}$,
 F.A.~Danevich$^{f}$,
 A.~Incicchitti$^{c,d}$,
 D.V.~Kasperovych$^{f}$,
 V.V.~Kobychev$^{f}$,
 G.P.~Kovtun$^{g,h}$,
 N.G.~Kovtun$^{g}$,
 M.~Laubenstein$^{e}$,
 D.V.~Poda$^{i}$,
 O.G.~Polischuk$^{f}$,
 A.P.~Shcherban$^{g}$,
 S.~Tessalina$^{j}$,
 V.I.~Tretyak$^{f}$
 }
 \vskip 0.3cm

$^{a}${\it INFN, sezione Roma ``Tor Vergata'', I-00133 Rome,
Italy}

$^{b}${\it  Dipartimento di Fisica, Universita di Roma ``Tor
Vergata, I-00133 Rome, Italy''}

$^{c}${\it  INFN, sezione Roma ``La Sapienza'', I-00185 Rome,
Italy}

$^{d}${\it  Dipartimento di Fisica, Universita di Roma ``La
Sapienza'', I-00185 Rome, Italy}

$^{e}${\it  INFN, Laboratori Nazionali del Gran Sasso, 67100
Assergi (AQ), Italy}

$^{f}${\it  Institute for Nuclear Research of NASU, 03028 Kyiv,
Ukraine}

$^{g}${\it  National Science Center ``Kharkiv Institute of Physics
and Technology'', 61108 Kharkiv, Ukraine}

$^{h}${\it  Karazin Kharkiv National University, 4, 61022 Kharkiv,
Ukraine}

$^{i}${\it Universit\'{e} Paris-Saclay, CNRS/IN2P3, IJCLab, 91405
Orsay, France}

$^{j}${\it John de Laeter Centre for Isotope Research, GPO Box U
1987, Curtin University, Bentley, WA, Australia}

\end{center}


\vskip 0.5cm

\begin{abstract}

A search for $\alpha$~decay of naturally occurring osmium isotopes
to the lowest excited levels of daughter nuclei has been
performed by using an ultra-low-background Broad-Energy Germanium
$\gamma$-detector with a volume of 112~cm$^3$ and an ultra-pure
osmium sample with a mass of 118 g at the Gran Sasso National
Laboratory of the INFN (Italy). The isotopic composition of the osmium sample has
been measured with high precision using Negative Thermal
Ionisation Mass Spectrometry. After 15851~h of data taking with
the $\gamma$-detector no effect has been detected, and lower limits
on the $\alpha$ decays were set at level of $\lim T_{1/2}\sim
10^{15}-10^{19}$~yr. The limits for the $\alpha$~decays of
$^{184}$Os and $^{186}$Os to the first excited levels of daughter
nuclei, $T_{1/2}(^{184}$Os$)\geq 6.8\times10^{15}$~yr and
$T_{1/2}(^{186}$Os$)\geq3.3\times10^{17}$~yr (at 90\% C.L.),
exceed the present theoretical estimates of the decays half-lives.
For $^{189}$Os and $^{192}$Os also
decays to the ground states of the daughter nuclei were searched
for due to the instability of the daughter nuclides relative to
$\beta$ decay.

\end{abstract}

\vskip 0.4cm


\section{Introduction}
\label{intro} The interest to $\alpha$ decay, a phenomenon
discovered more than 100 years ago \cite{rut99}, is still great,
both from the theoretical and the experimental sides. Various
theoretical models are continuously developed or improved (see
e.g.
\cite{den09,poe12,qia14,den15,san15,ash16,sah16,zha17,akr17,Tav20}
and references therein), in particular motivated by searches for
stable or long-lived super-heavy isotopes
\cite{hof15,oga17,Giuliani:2019} and predictions of their
half-lives.

Improvements in the experimental sensitivity, especially related
with the use of super-low-background set-ups located in
underground laboratories, have led during the last decade to the
discovery of $\alpha$ decays which were not observed previously
due to their extremely long half-lives. We refer the interested
readers to \cite{bel19} where the current status of the
experimental searches for rare $\alpha$ and $\beta$ decays is
reviewed. The half-life of $^{174}$Hf was remeasured recently
(after publication of the review \cite{bel19}) with improved
accuracy as $T_{1/2}=(7.0\pm1.2)\times10^{16}$ yr with the help of
a Cs$_2$HfCl$_6$ scintillator \cite{Caracciolo:2020}.

All the seven naturally occurring osmium isotopes are potentially
unstable relative to $\alpha$ decay (see Table~\ref{tab:theory}),
however, only for two of them (with the highest $Q_\alpha$ values)
indications on their existence were obtained. For $^{184}$Os, only
limits were known previously: $T_{1/2}>2.0\times10^{13}$ yr
\cite{por56,bag10} set with nuclear emulsions, and
$T_{1/2}>5.6\times10^{13}$ yr \cite{spe76} with proportional
counter measurements of Os sample enriched in $^{184}$Os to
2.25\%. However, recently an indication on $\alpha$ decay of
$^{184}$Os was found in geochemical measurements \cite{pet14}
where an excess of daughter $^{180}$W was measured in meteorites
and terrestrial rocks; the half-life was determined as
$T_{1/2}=(1.1\pm0.2)\times10^{13}$ yr, which is in contradictions
with the results of the direct laboratory measurements. The decay
of $^{186}$Os with $T_{1/2}=(2.0\pm1.1)\times10^{15}$ yr was
observed in direct experiment with a semiconductor detector and an
Os sample enriched in $^{186}$Os to 61.27\% \cite{vio75}.

\begin{table}
\scriptsize
 \caption{Characteristics of $\alpha$ decays of
naturally occurring osmium isotopes. $J^{\pi}$ is spin and parity
of the nuclei, $E$ is the energy of excited levels of the daughter
nuclei, the $Q_{\alpha}$ value is given for the g.s. to g.s.
transitions. The g.s. of the parent nuclei is assumed. The limits
obtained in the present work are given at 90\% confidence level
(C.L.).}
 \centering
 \label{tab:theory}
 \begin{tabular}{lllllll}
 \hline\noalign{\smallskip}
 Transition, $J^{\pi}$, $E$ (keV)                   & $Q_\alpha$ (keV)  & \multicolumn{5}{c}{Partial $T_{1/2}$ (yr)} \\
                                                    & \cite{wan17}      & \multicolumn{4}{c}{Theoretical}                                                   & Experimental \\
                                                    &                   & \cite{poe83}      &  \cite{Buck:1991,Buck:1992} & \cite{den15}          & \cite{Tav20}      &  \\
 ~                                                  &                   &                   &  ~                &  ~                    & ~                 &  \\
\noalign{\smallskip}\hline\noalign{\smallskip}
$^{184}$Os, $0^+~\rightarrow ^{180}$W, $0^+$, g.s.  & 2958.7(16)        & $7.2\times10^{13}$ & $3.5\times10^{13}$ & $3.3\times10^{13}$  & $2.1\times10^{13}$& $>2.0\times10^{13}$ \cite{por56}  \\
 ~                                                  &                   &~                  &~                  &~                      &~                  & $>5.6\times10^{13}$ \cite{spe76} \\
 ~                                                  &                   &~                  &~                  &~                      &~                  & $=(1.1\pm0.2)\times10^{13}$ \cite{pet14}\\
$^{184}$Os, $0^+\rightarrow ^{180}$W, $2^+$, 103.6  &                   &$2.9\times10^{15}$ & $1.3\times10^{15}$ & $6.3\times10^{14}$   & $7.3\times10^{14}$& $\geq 6.8\times10^{15}$ this work \\
$^{184}$Os, $0^+\rightarrow ^{180}$W, $4^+$, 337.6  &                   &$2.5\times10^{19}$ & $1.0\times10^{19}$ & $9.2\times10^{17}$   & $4.6\times10^{18}$& $\geq 4.6\times10^{16}$ this work  \\
 \noalign{\smallskip}
 $^{186}$Os, $0^+~\rightarrow ^{182}$W, $0^+$, g.s.  & 2821.2(9)        &$4.7\times10^{15}$ & $1.9\times10^{15}$ & $1.6\times10^{15}$   & $1.0\times10^{15}$& $=(2.0\pm1.1)\times10^{15}$ \cite{vio75}\\
 $^{186}$Os, $0^+~\rightarrow ^{182}$W, $2^+$, 100.1 &                  &$2.2\times10^{17}$ & $8.3\times10^{16}$ &$3.3\times10^{16}$    & $3.9\times10^{16}$ & $\geq 3.3\times10^{17}$ this work \\
 $^{186}$Os, $0^+~\rightarrow ^{182}$W, $4^+$, 329.4 & ~                &$2.9\times10^{21}$ & $9.7\times10^{20}$ &$7.2\times10^{19}$    & $3.7\times10^{20}$ & $\geq 6.0\times10^{18}$ this work \\
 \noalign{\smallskip}
 $^{187}$Os, $1/2^-~\rightarrow^{183}$W, $1/2^-$, g.s. & 2721.7(9)      &$4.5\times10^{19}$ & $4.1\times10^{16}$ &$5.1\times10^{16}$     &$2.0\times10^{16}$ & -- \\
 $^{187}$Os, $1/2^-~\rightarrow^{183}$W, $3/2^-$, 46.5 & ~              &$4.4\times10^{20}$ & $3.6\times10^{17}$ &$6.7\times10^{18}$     &$1.6\times10^{17}$ & $\geq 3.2\times10^{15}$ this work \\
 $^{187}$Os, $1/2^-~\rightarrow^{183}$W, $5/2^-$, 99.1 & ~              &$2.8\times10^{21}$& $2.1\times10^{18}$  &$4.0\times10^{19}$     &$9.1\times10^{17}$ & $\geq 1.9\times10^{17}$ this work \\
 \noalign{\smallskip}
 $^{188}$Os, $0^+~\rightarrow^{184}$W, $0^+$, g.s.  & 2143.2(9)         &$6.8\times10^{26}$ & $1.4\times10^{26}$ &$7.2\times10^{25}$     &$5.2\times10^{25}$ & --  \\
 $^{188}$Os, $0^+~\rightarrow^{184}$W, $2^+$, 111.2 &                   &$2.9\times10^{29}$ & $5.5\times10^{28}$ &$1.3\times10^{28}$     &~                  & $\geq 3.3\times10^{18}$ this work \\
 $^{188}$Os, $0^+~\rightarrow^{184}$W, $4^+$, 364.1 &                   &$1.9\times10^{36}$ & $2.7\times10^{35}$ &$8.9\times10^{33}$     &~                  & $\geq 5.0\times10^{19}$ this work \\
 \noalign{\smallskip}
 $^{189}$Os, $3/2^-~\rightarrow^{185}$W, $3/2^-$, g.s.& 1976.1(9)       &$2.4\times10^{34}$ & $4.8\times10^{29}$&$3.1\times10^{29}$     &~                  & $\geq 3.5\times10^{15}$ this work \\
 $^{189}$Os, $3/2^-~\rightarrow^{185}$W, $1/2^-$, 23.5& ~               &$1.8\times10^{35}$ & $3.2\times10^{30}$&$1.1\times10^{32}$     &~                  & $\geq 3.5\times10^{15}$ this work \\
 $^{189}$Os, $3/2^-~\rightarrow^{185}$W, $5/2^-$, 65.9& ~               &$2.1\times10^{36}$ & $3.1\times10^{31}$ &$1.1\times10^{33}$    &~                  & $\geq 7.6\times10^{17}$ this work \\
 \noalign{\smallskip}
 $^{190}$Os, $0^+~\rightarrow^{186}$W, $0^+$, g.s.  & 1375.8(12)        &$3.6\times10^{48}$ & $2.0\times10^{47}$ &$2.1\times10^{46}$    &~                  & -- \\
 $^{190}$Os, $0^+~\rightarrow^{186}$W, $2^+$, 122.6 & ~                 &$1.1\times10^{54}$ & $4.9\times10^{52}$&$1.6\times10^{51}$     &~                  & $\geq 1.2\times10^{19}$ this work \\
 $^{190}$Os, $0^+~\rightarrow^{186}$W, $4^+$, 396.5 & ~                 &$5.8\times10^{69}$ & $1.1\times10^{68}$ &$1.6\times10^{65}$    &~                  & $\geq 8.6\times10^{19}$ this work \\
 \noalign{\smallskip}
 $^{192}$Os, $0^+~\rightarrow^{188}$W, $0^+$, g.s.  & 361(4)             &$1.7\times10^{153}$& $1.8\times10^{149}$&$1.4\times10^{140}$  &~                  & $\geq 5.8\times10^{18}$ this work \\
 $^{192}$Os, $0^+~\rightarrow^{188}$W, $2^+$, 143.2 &                   &$1.6\times10^{215}$& $5.5\times10^{209}$ &$9.9\times10^{190}$  &~                  & $\geq 2.7\times10^{19}$ this work \\ \\
\noalign{\smallskip}\hline
\end{tabular}
\end{table}

\begin{figure}
\centering \resizebox{0.6\textwidth}{!}{
  \includegraphics{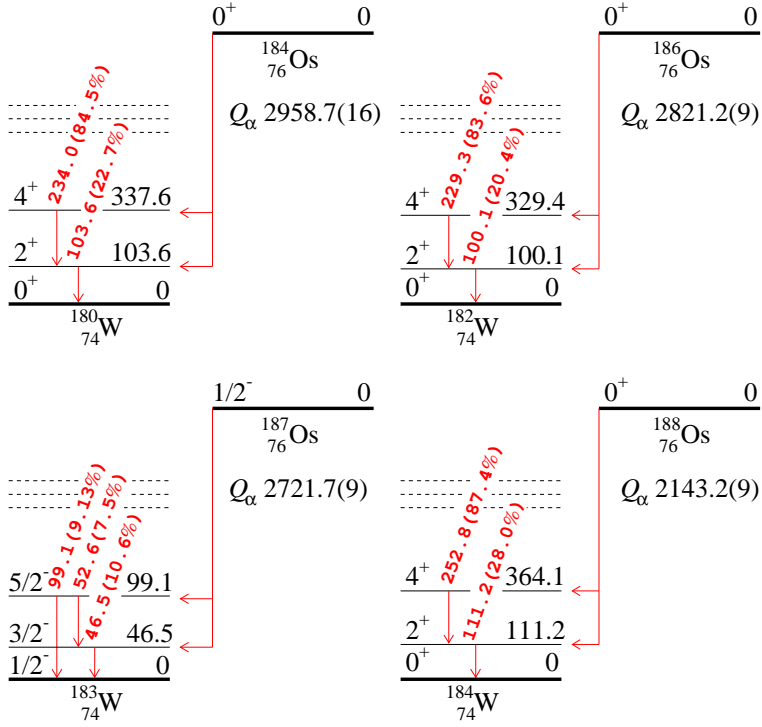}
} \caption{(Color online) Expected schemes of $^{184}$Os,
$^{186}$Os, $^{187}$Os and $^{188}$Os $\alpha$~decay to the two
first excited levels of daughter nuclei. The $Q_\alpha$ values,
energies of the levels and of the de-excitation $\gamma$ quanta
are given in keV; the probabilities of $\gamma$ quanta emission
are given in parentheses \cite{180W,182W,183W,bag10}.} \label{fig:decay_scheme1}
\end{figure}

\begin{figure}
\centering \resizebox{0.6\textwidth}{!}{
  \includegraphics{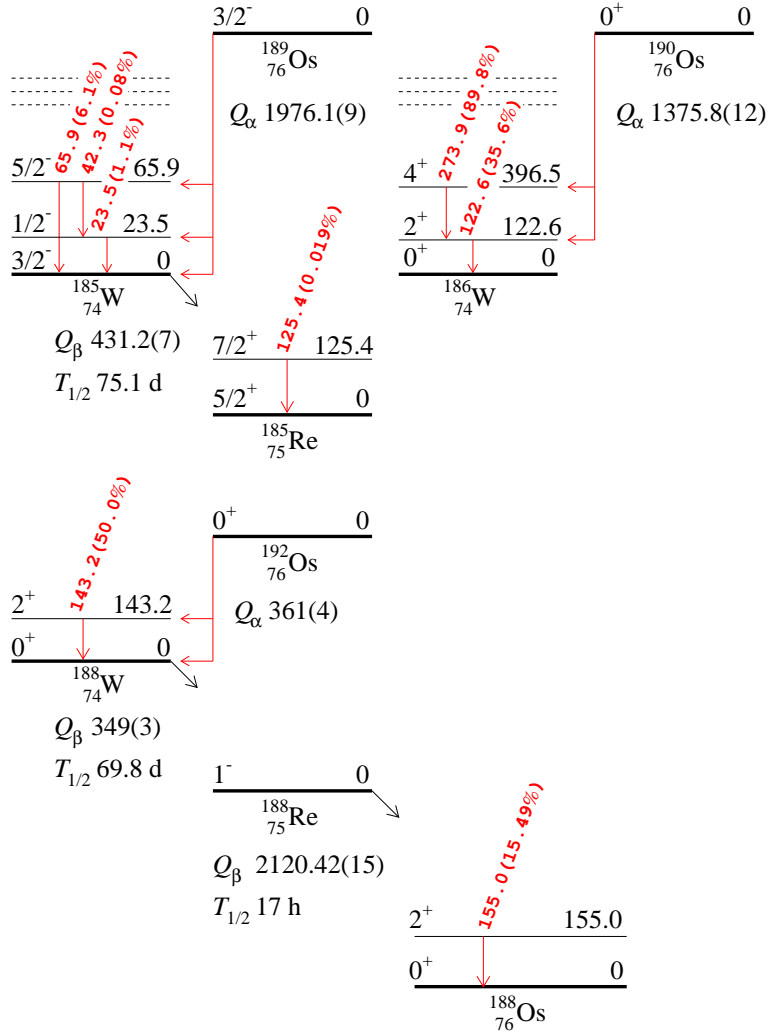}
} \caption{(Color online) Expected schemes of $^{189}$Os and
$^{190}$Os $\alpha$~decay to the two first excited levels of
  daughter nuclei. The decay scheme of $^{192}$Os is also reported. The $Q_\alpha$ values,
energies of the levels and of the de-excitation $\gamma$ quanta
are given in keV; the probabilities of $\gamma$ quanta emission
are given in parentheses \cite{185W_185Re,186W,188W_188Os}.} \label{fig:decay_scheme2}
\end{figure}

The process of $\alpha$ decay can be accompanied by the emission of
$\gamma$ quanta when the decay goes to excited level(s) of a daughter
nucleus. In this work, we look for $\gamma$ quanta expected in
$\alpha$ decays of the naturally occurring osmium nuclides to the
two lowest excited levels of daughter nuclei (see
Figs.~\ref{fig:decay_scheme1} and \ref{fig:decay_scheme2} where
expected schemes of $\alpha$~decay of the osmium isotopes are
shown). It should be noted that the $^{189}$Os and $^{192}$Os also
$\alpha$ decay to the ground state of the daughter nuclei can be
searched for thanks to the $\beta$-instability of the daughter
nuclides which is also accompanied by $\gamma$ quanta. The
experiment was realized with the help of ultra-low background HPGe
$\gamma$ spectrometry of a highly purified osmium metal sample
with the natural isotopic composition. The isotopic composition of
the osmium was measured precisely with the help of Negative Thermal
Ionization Mass Spectrometry. The results of the previous stage of
the experiment, which was devoted mainly to search for
$2\beta$~processes in $^{184}$Os and $^{192}$Os, were reported in
\cite{bel13a,bel13b}. The main attention in the present study is
focused on the search for $\alpha$ decays with emission of $\gamma$
quanta in $^{184}$Os and $^{186}$Os, taking into account the
theoretically highest decay probabilities for these nuclides (see
Table \ref{tab:theory}).

\section{Experiment}
\label{sec:experiment}

\subsection{Sample of osmium}
\label{sec:Os-sample}

Osmium in form of metal of at least 99.999\% purity grade
\cite{azha00} was used in the present experiment. The material was
obtained from osmium powder by electron-beam melting with further
purification by electron-beam zone refining at the National
Science Center ``Kharkiv Institute of Physics and Technology''
(Kharkiv, Ukraine). The osmium in form of four ingots with a total mass of 173 g 
was used in the first low-background experiment aiming at the
search for double beta processes in $^{184}$Os and $^{192}$Os
\cite{bel13a,bel13b}. Preliminary results of the searches for
$\alpha$ decay of $^{184}$Os and $^{186}$Os to the first excited
levels of daughter nuclei were reported too \cite{bel13b}.


The density of osmium metal is very high (in fact, osmium is the
densest naturally occurring element: the sample density was
estimated to be 23~g/cm$^3$, while the reference value is
22.587~g/cm$^3$ \cite{Haynes:2017}). Thus, $\gamma$-ray quanta
expected to be emitted in the $\alpha$ decays of the naturally
occurring osmium isotopes are strongly absorbed in the sample. To
increase the detection efficiency the ingots were cut into thin
slices with a thickness of ($0.79-1.25$) mm by using a method of
electroerosion cutting with brass wire in kerosene. The slices
were then etched in a solution of nitric and hydrochloric acids
and washed by ultra-pure water.

\subsection{Precise measurements of the osmium isotopic composition}
\label{sec:abundance}

The Os isotopic composition was analysed in the John de Laeter
Centre at Curtin University (Perth, Western Australia). In order
to achieve a complete digestion of the pure Os metal, the Carius
tube digestion method modified from Shirey and Walker
\cite{Shirey:1995} was applied. Approximately 0.5 mg of Os metal
was consumed for each of the two samples studied. The acid
digestion was done using concentrated acids (3 mL of purged
double-distilled HNO$_3$ and 1 mL of triple-distilled HCl). This
mixture was chilled and sealed in previously cleaned
Pyrex$^{\rm{TM}}$ borosilicate Carius Tubes and heated up to 220
$^{\circ}$C for 60 h. Osmium was extracted from the acid solution
by chloroform solvent extraction \cite{Cohen:1996}, then
back-extracted into HBr, followed by purification via
microdistillation \cite{Birck:1997}. The purified Os fraction of
each of the two samples was loaded onto five separate Pt filaments
(ten in total), and measured using Negative Thermal Ionisation
Mass Spectrometry (N-TIMS) on a Thermo-Fisher Triton$^{\rm{TM}}$
mass spectrometer using Faraday cup collectors. The beam for
samples studied was maintained at $\sim10^{-11}$ A (for
$^{192}$Os) for extended period of time for Os metal (100 blocks
of 10 cycles were collected), allowing to obtain a standard error
of the mean precision below 10 ppm level \cite{Birck:2001}. The
measured oxide ion ratios OsO$_3^-$ were corrected for isobaric
oxygen interferences to obtain element ratios, which were
corrected for mass fractionation using a $^{192}$Os/$^{188}$Os
value of 3.08271 \cite{Nier:1937}.

The pure Os metal gave the average of $^{187}$Os/$^{188}$Os ratio
$0.14179\pm0.00009$ at 95\% C.L. To monitor the intermediate
precision over a period of 12 months of the N-TIMS instrument for
Os, an AB-2 Os reference solution (University of Alberta) was
measured as part of the protocol. The AB-2 Os standard yielded the
$^{187}$Os/$^{188}$Os ratio $0.10687\pm0.00012$ (95\% C.L.) during
the 12 month period of the measurements, which is consistent with
that reported by Selby and Creaser \cite{Selby:2003}
$(0.10684\pm0.00004)$. The total procedural blank for Os was 0.50
pg, its contribution is insignificant for the sample studied. The
$^{187}$Os/$^{188}$Os ratio for the blank was $0.201\pm0.020$
($n=2$).

A summary of the osmium isotopic composition, as well as numbers
of nuclei of the osmium isotopes in the sample are given in Table
\ref{tab:abundance}. The accuracy of the $^{184}$Os measurement
[0.0170(7)\%] is similar to the accuracy of the ``best measurement
from a single terrestial source'' [0.0197(5)\%] \cite{volk91}, and
is definitely higher than the error scale recommended by IUPAC:
0.02(2)\% \cite{mei16}. The $^{186}$Os isotopic concentration was
measured with a much higher accuracy than the recommended
representative isotopic abundance too. Also for other Os isotopes
the errors in the present study are much smaller than the ones
given in \cite{mei16}\footnote{Moreover, there is still a room for
improvement of the Os isotopic composition accuracy as it was
recently demonstrated in \cite{Zhu:2018}.}.

\begin{table}
\caption{Isotopic composition ($\delta$) of the osmium measured in
the present work and the numbers of nuclei of each isotope in the
sample calculated by using the measured isotopic concentrations.
The representative isotopic abundances from \cite{mei16} are given
too.}
 \centering
\label{tab:abundance}
\begin{tabular}{cccc}
\hline\noalign{\smallskip}

 Isotope            & \multicolumn{2}{c}{$\delta$ (\%)} & Number of nuclei\\
                    & IUPAC \cite{mei16}  & this work    & in the sample   \\
\noalign{\smallskip}\hline\noalign{\smallskip}

$^{184}$Os    & 0.02(2)       & 0.0170(7)   & $6.35(26)\times10^{19}$ \\
~                   &               & ~            & ~                       \\
$^{186}$Os    & 1.59(64)      & 1.5908(6)   & $5.9405(25)\times10^{21}$ \\
~                   &               & ~            & ~                       \\
$^{187}$Os    & 1.96(17)      & 1.8794(6)   & $7.0182(25)\times10^{21}$ \\
~                   &               & ~            & ~                       \\
$^{188}$Os    & 13.24(27)     & 13.253(3)   & $4.9490(14)\times10^{22}$ \\
~                   &               & ~            & ~                       \\
$^{189}$Os    & 16.15(23)     & 16.152(4)   & $6.0316(18)\times10^{22}$ \\
~                   &               & ~            & ~                       \\
$^{190}$Os    & 26.26(20)     & 26.250(8)   & $9.8025(34)\times10^{22}$ \\
~                   &               & ~            & ~                       \\
$^{192}$Os    & 40.78(32)     & 40.86(5)    & $1.5258(19)\times10^{23}$ \\

\noalign{\smallskip}\hline
\end{tabular}
\end{table}

\subsection{Low-background measurements}
\label{sec:measurements}

\begin{figure}[htbp]
\begin{center}
\resizebox{0.5\textwidth}{!}{\includegraphics{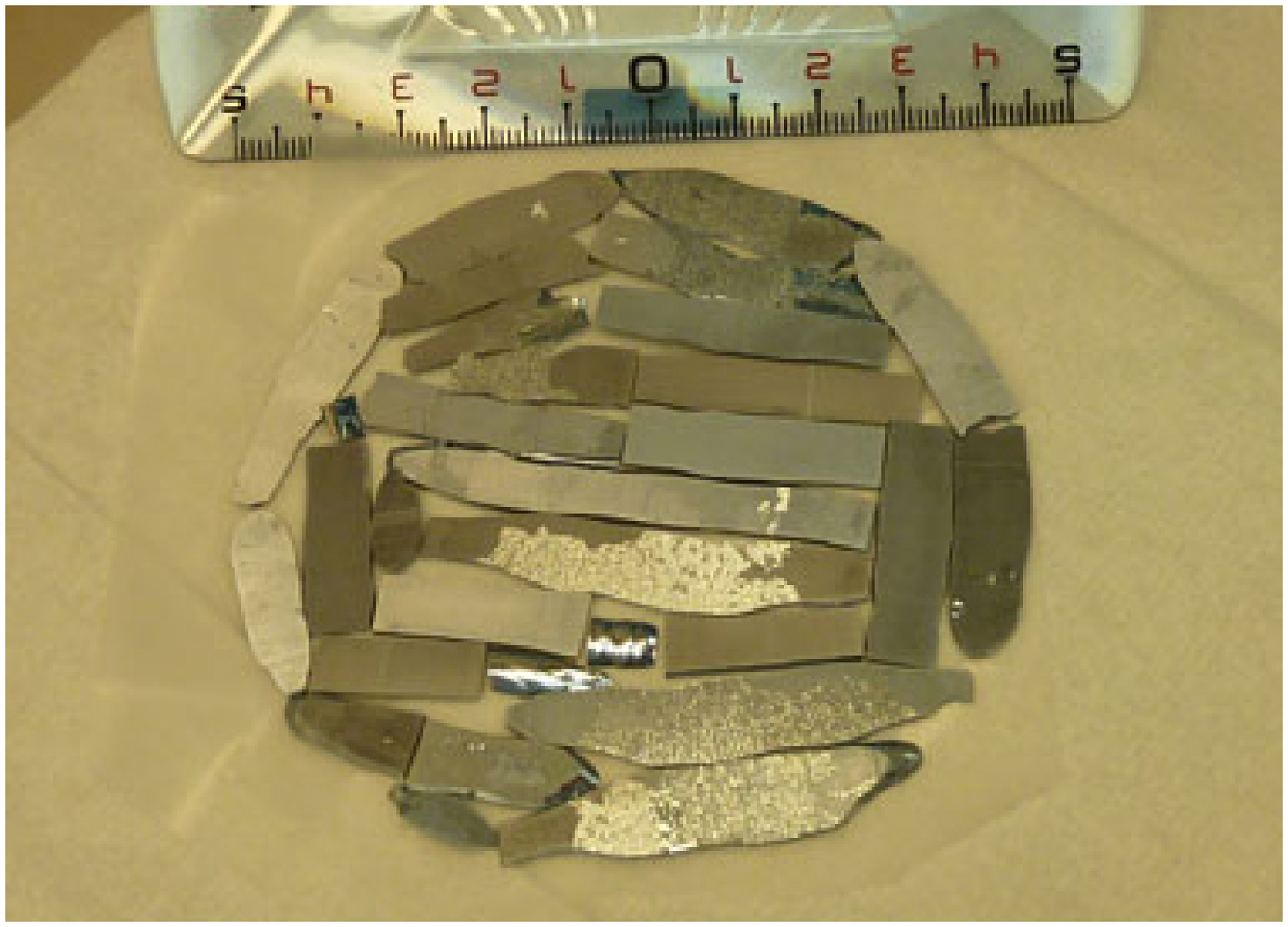}}
\resizebox{0.455\textwidth}{!}{\includegraphics{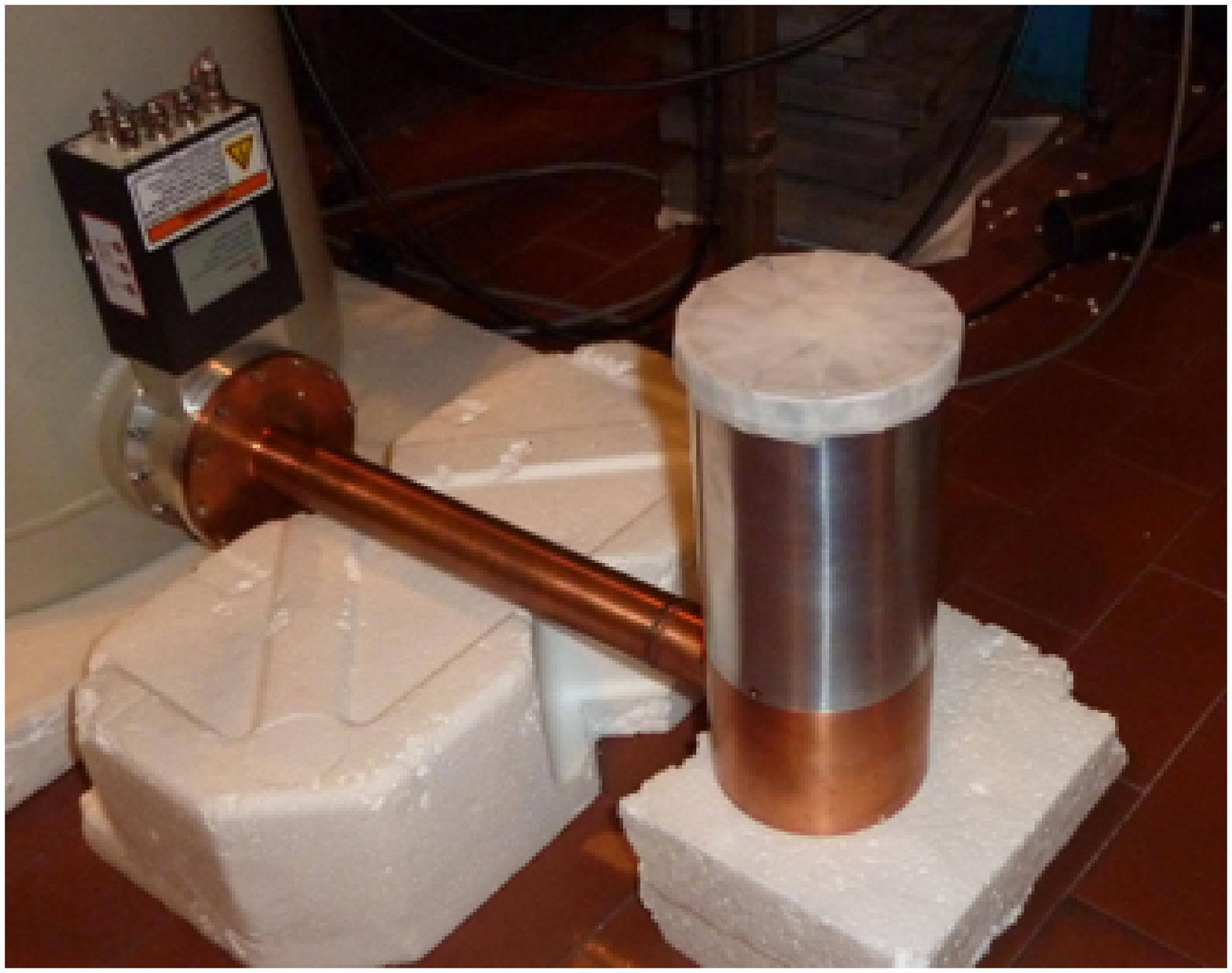}}
\caption{(Color online) Left photograph: the main part of the
osmium slices before assembling on a plastic Petri dish (the scale
is in centimeters). Right photograph: arrangement of the sample on
a cryostat end-cap (unshielded detector, not the same used in the measurements).}
\end{center}
 \label{fig:photo}
\end{figure}

The Os slices (see Sec. \ref{sec:Os-sample}) with a total mass of
117.96(2)~g were fixed on the inner surface of a plastic Petri
dish (with a thickness of 0.8 mm) with the help of Scotch 811
removable tape. The Petri dish with the Os slices was installed
directly on the aluminium end-cap of the cryostat of the ultra-low
background Broad-Energy Germanium (BEGe) detector with a volume of
112.5~cm$^3$ (Fig.~\ref{fig:photo}). The detector, thanks to a
very thin dead layer of 0.4 $\mu$m, offers a high sensitivity to
low-energy photons. The detector with the Os sample was shielded
by layers of $\approx5$ cm thick copper and 20 cm thick lead.

\nopagebreak

The low-background measurements were carried out at the STELLA
(SubTerranean Low Level Assay) facility of the Gran Sasso National
Laboratory of the INFN (Italy) \cite{matthias}. The laboratory is located at a
depth of $\approx3600$ meters of water equivalent. The data were taken
in 10 runs with a total measurement time of 15851~h. The energy
scale of the detector was measured with $\gamma$ sources in the
beginning of the experiment. Then the data of each run were
re-calibrated by using clear and intensive background $\gamma$
peaks of $^{40}$K, $^{208}$Tl, $^{210}$Pb, $^{214}$Pb and
$^{214}$Bi to improve the energy resolution in the final sum
spectrum. The dependence of energy resolution (full width at half
maximum, FWHM) on energy of $\gamma$-ray quanta ($E_{\gamma}$, in
keV) in the sum energy spectrum can be approximated by the
following function:

\begin{equation}\label{eq:fwhm}
\rm{FWHM~(keV)}=0.57(5)+0.029(2)\times\sqrt{\textit{E}_{\gamma}}.
\end{equation}

\noindent 
The re-calibration procedure allowed to improve the detector
energy resolution in the final spectrum by 13\% (at energy 100
keV) in comparison to the sum energy spectrum obtained without the
correction.


\section{Results and discussion}
\label{sec:Res-Disc}

\subsection{Radioactive contamination of the osmium sample}

\begin{figure}[hb]
\centering
 \mbox{\epsfig{figure=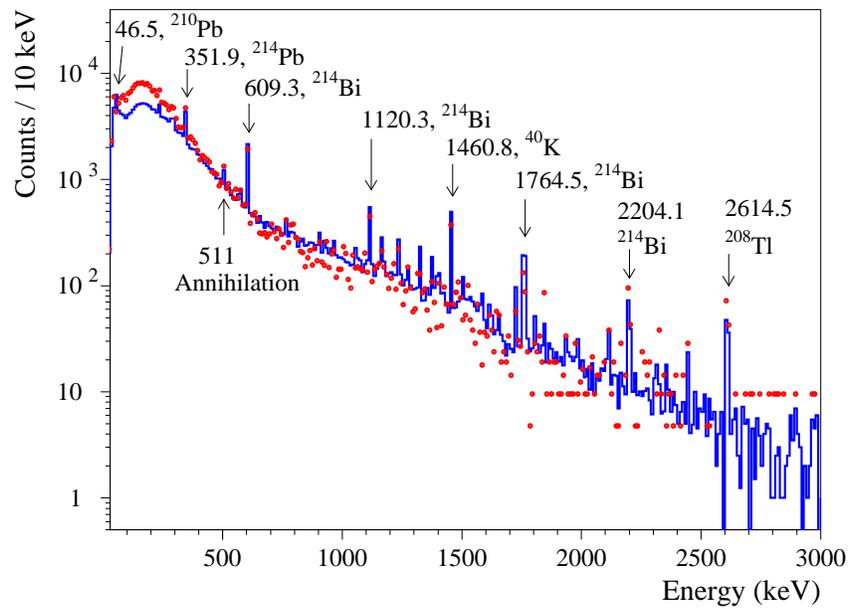,height=8.cm}}
\caption{Energy spectra accumulated for 15851~h with the Os
sample (solid line, blue online) and background energy spectrum
measured over 1660~h without sample (dots; red online). The
background data are normalized to the time of measurements with
the Os sample. Energies of $\gamma$-ray peaks are in keV.}
 \label{fig:spectrum}
\end{figure}

The energy spectrum measured with the Os sample for 15851~h is
shown in Fig.~\ref{fig:spectrum} together with the background
energy spectrum taken over 1660~h. One can see that the counting
rate in the spectrum measured with the Os sample below $\approx
0.4$ MeV is lower than in the background data. The difference is
due to the very high density of osmium. As a result, the Os sample
is effectively absorbing radiations from the shielding materials
around the detector (mainly bremsstrahlung $\gamma$-ray from
$^{210}$Pb in the lead details of the shielding), and thus
reducing the count rate at low energies.

\clearpage

 The peaks in the energy spectra belong mainly to
$\gamma$-ray quanta of naturally occurring primordial
radionuclides: $^{40}$K, and  daughters of $^{232}$Th, $^{235}$U,
$^{238}$U. There are also weak peaks in both the spectra that can
be ascribed to $^{60}$Co and $^{137}$Cs. Specific activities of
the radionuclides in the Os sample were calculated with the
following formula:

\begin{equation}
A = (S_{sample}/t_{sample}-S_{bg}/t_{bg})/(\eta \cdot \varepsilon
\cdot m),
 \end{equation}

\noindent where $S_{sample}$ ($S_{bg}$) is the area of a peak in
the sample (background) spectrum; $t_{sample}$ ($t_{bg}$) is the
time of the sample (background) measurement; $\eta$ is the
$\gamma$-ray emission intensity; $\varepsilon$ is the full energy
peak efficiency; $m$ is the sample mass. The detection
efficiencies to $\gamma$-ray quanta were calculated using the
GEANT4 simulation package
\cite{Agostinelli:2003,Allison:2006,Boswell:2011}, the decay
events were generated homogeneously in the Os sample. If no
statistically significant peak excess was observed (the cases of
$^{60}$Co, daughters of $^{232}$Th, $^{235}$U and $^{238}$U), only
upper limits on the specific activities of the radioactive
impurities in the sample were set.
A summary of the Os sample radioactive contamination is given in
Table 3.

\nopagebreak
\begin{table}[htb]
\label{tab:rad-cont}
 \caption{Radioactive contamination of the Os
sample measured by the ultra-low-background BEGe $\gamma$
detector.}
\begin{center}
\begin{tabular}{|l|l|l|l|}
 \hline
 Decay chain    & Radionuclide  & Specific activity (mBq/kg) \\
 \hline
 ~              & $^{40}$K      & $11\pm4$ \\
 ~              & $^{60}$Co     & $\leq1.3$ \\
 ~              & $^{137}$Cs    & $0.5\pm0.1$ \\
 \hline
 $^{232}$Th     & $^{228}$Ra    & $\leq6.6$ \\
 ~              & $^{228}$Th    & $\leq16$ \\
 \hline
 $^{235}$U      & $^{235}$U     & $\leq8.0$ \\
 ~              & $^{231}$Pa    & $\leq3.5$ \\
 ~              & $^{227}$Ac    & $\leq1.1$ \\
 \hline
 $^{238}$U      & $^{238}$U     & $\leq35$ \\
 ~              & $^{226}$Ra    & $\leq4.4$ \\
 ~              & $^{210}$Pb    & $\leq180$ \\
\hline
\end{tabular}
\end{center}
\end{table}

\subsection{Limits on $\alpha$ decays of $^{184}$Os and $^{186}$Os to the first
excited levels of daughter nuclei}

There are no peaks in the energy spectrum measured with the Os
sample that can be interpreted as $\alpha$ decay of naturally
occurring osmium nuclides. Thus, by analysis of the data one can
set half-life limits on the processes searched for with the help
of the following formula:

\begin{equation}
 \lim T_{1/2}=\frac{N\cdot \ln 2\cdot \eta \cdot \varepsilon \cdot t}{\lim S},
 \label{eq:t1/2}
\end{equation}

\noindent where \textit{N} is the number of nuclei of the isotope of
interest (given in Table \ref{tab:abundance}), $\eta$ is the
$\gamma$ quanta emission intensity (see Figs.
\ref{fig:decay_scheme1} and \ref{fig:decay_scheme2}),
$\varepsilon$~is the detection efficiency for $\gamma$~quanta
expected in the decays, $t$ is the time of measurement (15851~h),
and $\lim S$ is the upper limit on the number of events of the
effect searched for which can be excluded at a given C.L.

The detection efficiencies to $\gamma$-ray quanta expected in the
decays searched for were Monte Carlo simulated with the GEANT4
\cite{Agostinelli:2003,Allison:2006,Boswell:2011} and the EGSnrc
\cite{EGS} packages in two geometries: with uniform and granulated
source. In the ``uniform geometry'' the source was approximated by
a disc of 88 mm in diameter with a thickness of 0.88 mm, plus a
ring with an inner diameter of 90 mm and a height of 8 mm, with
the same thickness. The ``granulated geometry'' reproduces the
actual geometry of the source in a more accurate way (separate
objects with gaps between them). Both the GEANT4 and EGSnrc codes
give compatible results with the standard deviation of the
relative difference $2.7\%$ for $\gamma$ quanta in the energy
range from 46.5 keV to 273.0 keV (for the ``uniform geometry'').
The main difference in the simulations results is due to the
different source geometries. The relative difference between the
detection efficiencies for the ``uniform'' and ``granulated''
geometries decreases from 11.9\% to 4.5\% with increase of
$\gamma$ quanta energy from 46.5 keV to 273.0 keV, with a
systematically higher efficiency for the granulated source. The
higher efficiency for the ``granulated geometry'' can be explained
by the contribution of $\gamma$-quanta events emitted from the
side parts of the osmium slices (in contrary to the uniform source
geometry with no gaps in the Os material). The increase of the
difference with decrease of the $\gamma$-quanta energy can be
explained by the ``edge effect'' that is more significant at low
energies. Taking into account that the ``granulated geometry''
describes the sample in a more accurate way we use the detection
efficiencies obtained with this geometry for the further analysis.
The full energy peak detection efficiencies for the $\alpha$
decays under study are presented in Table \ref{tab:eff}.

\begin{table*}[ht]
\caption{Full energy peak detection efficiencies, $\varepsilon$, for signature
$\gamma$-ray quanta with energy $E_{\gamma}$, $\gamma$ quanta
emission intensity $\eta$, measured numbers of events ($S$), their
standard deviations ($\Delta S$) and estimated values of $\lim S$ (see discussions in the text)
for $\alpha$ transitions with emission of $\gamma$ quanta in
naturally occurring osmium isotopes. The relative systematic
uncertainties $\sigma_{\rm{r}}$ and factors $a$ (see text below)
to take into account the systematic uncertainties are given in the
last two columns.}
\begin{center}
\resizebox{\textwidth}{!}{
\begin{tabular}{|l|l|l|l|l|l|l|l|l|}
 \hline
 Transition                                         & $E_{\gamma}$  & $\eta$    & $\varepsilon$ & $S$   & $\Delta S$& $\lim S$  & $\sigma_{\rm{r}}$ & $a$ \\
                                                    & (keV)         &           &               &       &           &           &                   & \\
 \hline
 $^{184}$Os, $0^+\rightarrow ^{180}$W, $2^+$, 103.6 & 103.6         & 0.227      & 0.01382       & $-7.3$& 21.3      & 28.1      & 0.130             & 1.301 \\
 \hline
 $^{184}$Os, $0^+\rightarrow ^{180}$W, $4^+$, 337.6 & 234.0         & 0.845      & 0.05097       & 22.4  & 24.6      & 62.7      & 0.095            & 1.183 \\
 \hline
 $^{186}$Os, $0^+\rightarrow ^{182}$W, $2^+$, 100.1 & 100.1         & 0.204      & 0.01274       & 8.6   & 21.8      &  44.4     & 0.135             & 1.325 \\
 \hline
 $^{186}$Os, $0^+\rightarrow ^{182}$W, $4^+$, 329.4 & 229.3         & 0.836      & 0.05090       &$-14.2$& 24.9      &  28.0     & 0.206             & 1.893 \\
 \hline
 $^{187}$Os, $1/2^-\rightarrow^{183}$W, $3/2^-$, 46.5 & 46.5        & 0.106      & 0.00695       & 1272  & 42        &  1341     & 0.120             & 1.493 \\
 \hline
 $^{187}$Os, $1/2^-\rightarrow^{183}$W, $5/2^-$, 99.1 & 99.1        & 0.0913     & 0.01219       & 2.9   & 22.8      &  40.3     & 0.117             & 1.254  \\
 \hline
 $^{188}$Os, $0^+\rightarrow^{184}$W, $2^+$, 111.2  & 111.2         & 0.280      & 0.01629       & 3.8   & 30.7      &  54.1     & 0.150             & 1.568 \\
 \hline
 $^{188}$Os, $0^+\rightarrow^{184}$W, $4^+$, 364.1  & 252.8         & 0.874      & 0.05200       & $-2.9$& 25.8      &  39.7     & 0.140             & 1.420 \\
 \hline
 $^{189}$Os, $3/2^-\rightarrow^{185}$W, $3/2^-$, g.s.& 125.4        & 0.00019    & 0.02098       & $-2.7$& 24.9      & 38.4      & 0.245             & 2.233 \\
 \hline
 $^{189}$Os, $3/2^-\rightarrow^{185}$W, $5/2^-$, 65.9& 65.9         & 0.061      & 0.01936       & 30.7  & 23.6      & 69.4      & 0.189             & 1.689 \\
 \hline
 $^{190}$Os, $0^+\rightarrow^{186}$W, $2^+$, 122.6 & 122.6          & 0.356      & 0.02085       & 17.3  & 24.4      & 57.3      & 0.122             & 1.296 \\
 \hline
 $^{190}$Os, $0^+\rightarrow^{186}$W, $4^+$, 396.5 & 273.9          & 0.898      & 0.05197       & 21.3  & 24.3      & 61.2      & 0.067             & 1.090 \\
 \hline
 $^{192}$Os, $0^+\rightarrow^{188}$W, $0^+$, g.s.  & 155.0          & 0.1549     & 0.03220       & 93.7  & 29.3      & 142       & 0.081             & 1.159 \\
 \hline
 $^{192}$Os, $0^+\rightarrow^{188}$W, $2^+$, 143.2 & 143.2          & 0.500      & 0.02898       & 42.5  & 27.0      & 86.8      & 0.090             & 1.180 \\
  \hline
\end{tabular}
}
\label{tab:eff}
\end{center}
\end{table*}

\begin{figure}
\centering
 \mbox{\epsfig{figure=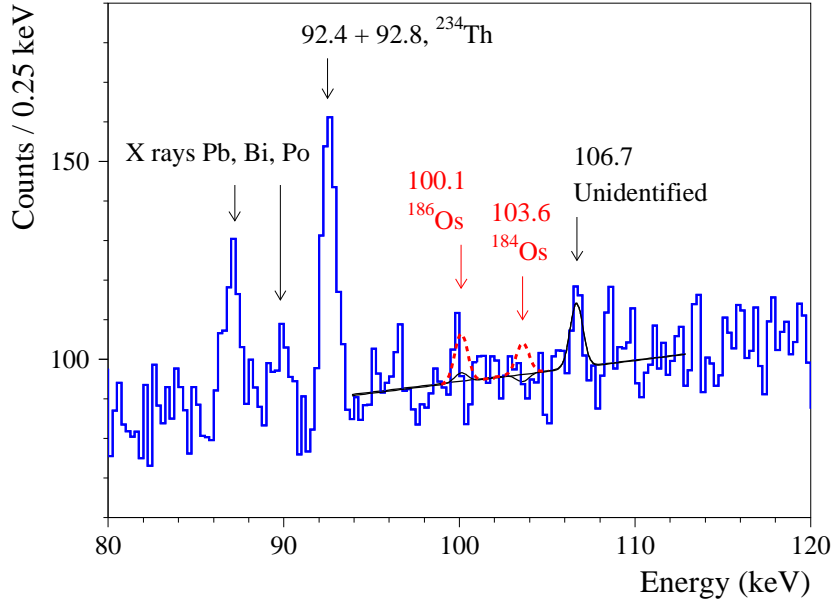,height=8.cm}}
\caption{Energy spectrum measured with the Os sample for 15851~h
in the region where the peaks with energies 100.1 keV and 103.6 keV
after the $\alpha$ decay of the $^{186}$Os and of the $^{184}$Os to the first
excited levels of daughter nuclei are expected. The fits of the data
by the background model (see text) are shown by solid lines (the
fits for the 100.1 keV and 103.6 keV peaks are almost
indistinguishable). The peaks with energies 100.1 keV and 103.6
keV excluded at 90\% C.L. are shown by dashed lines.}
 \label{fig:fit}
\end{figure}

To estimate the values of $\lim S$, the energy spectrum taken with the
Os sample was fitted in the region of interest for a certain
transition with the models accounting for the effect searched for
and for the background. For instance, to estimate value of $\lim
S$ for the decay of the $^{184}$Os to the first excited level of
$^{180}$W, the energy spectrum was fitted by a background model
constructed from a linear function (to describe the continuous
distribution), a peak with the energy of 103.6 keV (the effect
searched for), and an unidentified peak with energy
$\approx107$~keV. The data were fitted by the model with 5 free
parameters: two parameters of the linear function, an area of the
103.6 keV peak, an area and a position of the unidentified peak.
The peaks widths were fixed according to the estimated energy
dependence of the detector energy resolution (see 
formula given in eq. (\ref{eq:fwhm})). The best fit was
achieved in the energy interval $(93.75-112.75)$~keV with
$\chi^2=33.9$ for 72 degrees of freedom. The Fig.~\ref{fig:fit}
shows the energy spectrum measured with the Os sample in the
region of interest and its fit by the above described model. A
similar analysis was performed also for the 100.1 keV peak
expected in the $\alpha$ decay of $^{186}$Os to the first excited
level of $^{182}$W.

The fits return the 100.1 keV and 103.6 keV peaks areas
$(8.6\pm21.8)$~counts and $(-7.3\pm21.3)$~counts, respectively,
that is no evidence of the effects searched for\footnote{The area
of the unidentified peak is ($60\pm23$) counts, the energy of the
peak is ($106.67\pm0.15$) keV.}. By using the recommendations
\cite{feld98} we get the following upper limits on the peaks
areas: $\lim S=44.4$ counts (for the 100.1 keV peak), and $\lim
S=28.1$ counts (103.6 keV) at 90\%~C.L. (the excluded peaks are
shown in Fig. \ref{fig:fit}), that correspond to the lower
half-life limits $T_{1/2}(^{184}\rm{Os})>8.9\times10^{15}$~yr and
$T_{1/2}(^{186}\rm{Os})>4.4\times10^{17}$~yr, respectively. However, the limits
include only statistical errors.

Possible sources of systematic uncertainties of the half-life
limits are listed in Table \ref{tab:syst}. The systematic
uncertainties of the detection efficiencies were conservatively
estimated as the relative differences between the GEANT4 simulations
results with the ``granulated'' and ``uniform'' geometries.
Variations of $\lim S$ depending on the fit interval were
estimated by analysis of $\lim S$ distributions obtained from the
fit of the data in the energy intervals within $(93.5-96)$ keV for
the starting point, and $(110-116)$ keV for the final point, with
a step of 0.25 keV. An impact of the isotopic abundance
measurements uncertainties (presented in Table
\ref{tab:abundance}) was taken in consideration too. The total
relative systematic errors $\sigma_{\rm{r}}$ were obtained by
adding all the systematic contributions in quadrature.

\begin{table}[ht]
\caption{Estimated relative systematic uncertainties of the $^{184}$Os and
$^{186}$Os half-life limits relative to $\alpha$ decays to the
first excited levels of the daughter nuclei.}
\begin{center}
\begin{tabular}{|l|l|l|}
 \hline
 Source                 & \multicolumn{2}{c|}{Nuclide } \\
\cline{2-3}
  ~                     & $^{184}$Os    & $^{186}$Os \\
 \hline
 Detection efficiency   & 0.098           &  0.118 \\
 \hline
 Interval of fit        & 0.076           &  0.065 \\
 \hline
 Isotopic abundance     & 0.041           &  0.0004 \\
 \hline
 Total relative systematic error ($\sigma_{\rm{r}}$) & 0.131   & 0.135 \\
 \hline
\end{tabular}
\label{tab:syst}
\end{center}
\end{table}

The systematic uncertainties $\sigma_{\rm{r}}$ can be introduced
into the obtained lower half-life limits by correction of the
upper limits on the number of excluded events:

\begin{equation}
 \lim S^{'}=\lim S \times a,
\end{equation}

\noindent where $ \lim S^{'}$ is a corrected upper limit, and the
factor $a$ is expressed by the formula proposed in
\cite{Cousins:1992}:

\begin{equation}
 a=[1+(\lim S - S)\times \sigma_{\rm{r}}^{2}/2].
\end{equation}

After the correction the following half-life limits of $^{184}$Os
and $^{186}$Os relative to $\alpha$ decay to the first excited
levels of the daughter nuclei were obtained:

\begin{center}

$T_{1/2}(^{184}\rm{Os})>6.8\times10^{15}$~yr,

\vskip 0.3cm

$T_{1/2}(^{186}\rm{Os})>3.3\times10^{17}$~yr.

\end{center}

It should be noted that the limits substantially exceed the present theoretical
predictions (see Table \ref{tab:theory}). In
particular, the limits are one order of magnitude higher than the
estimates obtained by using the empirical relationships based on
the unified model for $\alpha$ decay and $\alpha$ capture (UMADAC)
\cite{den15} and the half-life values calculated in the framework
of a semi-empirical model based on the quantum mechanical
tunneling mechanism through a potential barrier \cite{Tav20}.

\subsection{Limits on other $\alpha$ decays of osmium nuclides with emission of $\gamma$ quanta}

Due to the smaller energy release, the theoretical predictions on
half-lives of $\alpha$ decays of $^{184}$Os and $^{186}$Os to the
second excited levels of daughter nuclei, not to say for $\alpha$
decays of other osmium nuclides, are much longer. Thus, the
sensitivity of the present experiment looks too low to detect the
processes. Nevertheless, the experimental data were used to
analyze other possible decays of the osmium nuclides too.

\begin{figure}[ht]
\centering
 \mbox{\epsfig{figure=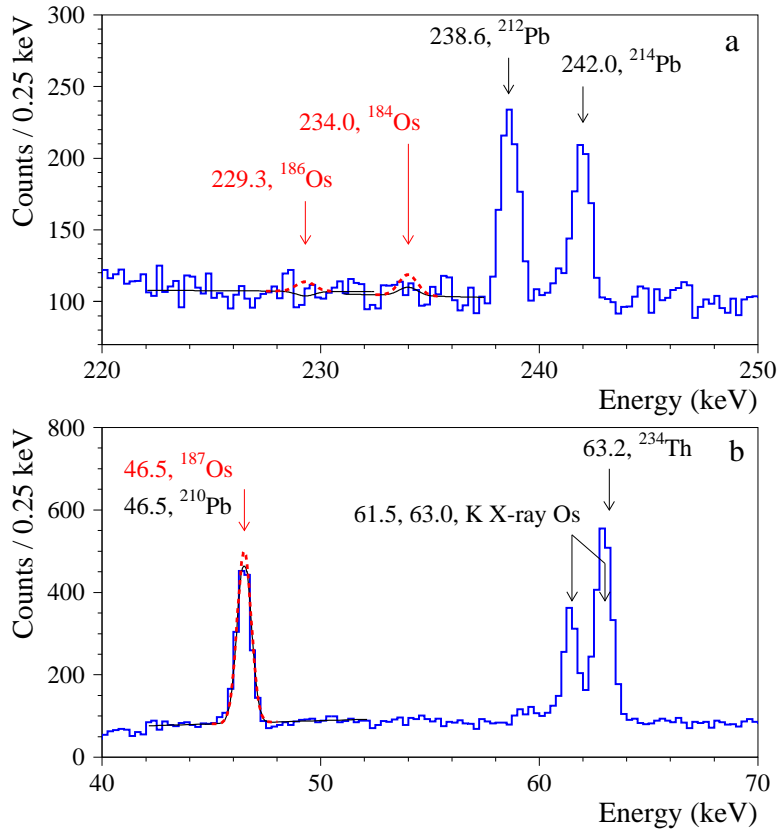,height=11.cm}}
\caption{Energy spectrum measured with the Os sample for 15851~h
in the region where peaks with energies 234.0 keV ($\alpha$ decay
of $^{184}$Os to the 337.6 keV excited level of $^{180}$W) and
229.3 keV ($\alpha$ decay of $^{186}$Os to the 329.4 keV excited
level of $^{182}$W) are expected (a). Low energy part of the
spectrum, where a peak with an energy of 46.5 keV is expected
($\alpha$ decay of $^{187}$Os to the 46.5 keV excited level of
$^{183}$W) (b). The existing peak with this energy can be
explained by $\gamma$ quanta after $\beta$ decay of $^{210}$Pb.
Fits of the data by the background model are shown by solid lines.
The peaks searched for, excluded at 90\% C.L., are shown by dashed
lines.}
 \label{fig:fit2}
\end{figure}

Examples of the energy spectrum fits in the regions of interest
for the $\alpha$ decays of $^{184}$Os and $^{186}$Os to the second
excited levels of $^{180}$W and $^{182}$W, and for the $\alpha$
decays of $^{187}$Os to the first 46.5 keV excited level of
$^{183}$W are shown in Fig. \ref{fig:fit2}. Unfortunately, the
signature $\gamma$ peak expected in the decay of $^{187}$Os
interferences strongly with the 46.5 keV $\gamma$-ray peak of
$^{210}$Pb (daughter of $^{222}$Rn from the $^{238}$U family),
typically present in low-background $\gamma$ spectra. Thus, we
accept the peak area (1272 counts) plus its standard error (42
counts) multiplied by 1.64 to get an estimation of $\lim S$ for
this decay channel at 90\% C.L.

To estimate the $\lim S$ values for other possible decay channels the
experimental spectrum was analyzed in the different energy
intervals in a similar way as described above. The detection
efficiencies and the emission intensities for the signature
$\gamma$-ray quanta, the obtained values of $S$, $\Delta S$ and
$\lim S$, the relative systematic errors $\sigma_{\rm{r}}$ and the
factors $a$ to account systematic uncertainties of the limits are
given in Table \ref{tab:eff}, while the half-life limits on the
$\alpha$ decays of naturally occurring osmium isotopes with
emission of $\gamma$ quanta are summarized in Table
\ref{tab:theory}.

It should be noted that for $^{189}$Os and $^{192}$Os also decays
to the ground states of the daughter nuclei (in general, to any
states of the daughter nuclei) were set due to the
$\beta$-instability of the daughter nuclides with lifetimes short
enough to be in equilibrium with the parent nuclides. However, the
theoretically estimated half-lives of the nuclides are very long
to be observed in a realistic experiment. At the same time our
experiment is not sensitive to the $\alpha$ decay of $^{189}$Os to
the first 23.5 keV excited level of $^{185}$W due to an
approximately two times higher energy threshold of the detector
$\approx42$ keV. Nevertheless, the limit obtained for the g.s. to
g.s. transition (by using the $\beta$-instability of $^{185}$W) is
valid for $\alpha$ decay to all levels, including excited ones (it
should be stressed, however, that also this decay is expected to
be too rare to be observed).

It should be also noted that a significant area ($93.7\pm29.3$)
counts of the 155.0 keV peak (expected in the decay sequence
$^{192}$Os$~\rightarrow~$$^{188}$W$~\rightarrow~$$^{188}$Re$~\rightarrow~$$^{188}$Os)
was interpreted as absence of the effect searched for. The peak
can be explained by several sources:

\begin{itemize}

\item $\gamma$-ray quanta with energy 154.0 keV of $^{228}$Ac
(daughter of $^{228}$Ra from the $^{232}$Th family; emission
probability 0.722\%).

\item $\gamma$-ray quanta with energy 154.2 keV of $^{223}$Ra
(daughter of $^{227}$Ac from the $^{235}$U family; emission
probability 5.7\%).

\item $\gamma$-ray quanta with energy 155.0 keV of $^{188}$Re
(daughter of $^{188}$W that can be cosmogenically-produced in
osmium; emission probability 15.49\%).

\item Thermal neutron capture $\gamma$-ray quanta with energy
155.0 keV emitted with an intensity 65\% after neutron-captures in
$^{187}$Os that has a rather big thermal neutron cross section
($320\pm10$ barn).

\end{itemize}

 \noindent
Precise determination of the 155.0 keV peak area is difficult
since the thermal neutrons flux in the set-up, activity and
localization of the $^{228}$Ra and $^{227}$Ac are unknown. Also
estimation of cosmogenic $^{188}$W activity in the Os sample is
not a trivial task. Here we conservatively ascribe all the counts
in the 155 keV peak (with error bar) to the $^{192}$Os alpha
decay.

\section{Theoretical estimates of the half-lives}

We calculated theoretical half-lives of Os nuclides relative to
$\alpha$ decay using the semi-empirical formulae \cite{poe83}
based on the liquid drop model and the description of $\alpha$
decay as a very asymmetric fission process. As well, the cluster
model of Refs. \cite{Buck:1991,Buck:1992} was used. The approaches
\cite{poe83,Buck:1991,Buck:1992} were tested with a set of
experimental half-lives of almost four hundred $\alpha$ emitters
and demonstrated good agreement between calculated and
experimental $T_{1/2}$ values, mainly inside a factor of $2-3$.
For Os $\alpha$ decays with a difference between spins and
parities of the parent and the daughter nuclei, which resulted in
non-zero angular momentum $l$ of the emitted $\alpha$ particle, we
take into account the additional hindrance factor $HF$, calculated
in accordance with \cite{Hey99} (for the lowest possible $l$
value). In particular, for the most important $\alpha$ decays of
$^{184}$Os and $^{186}$Os to the first excited levels of the
daughter nuclei ($0^+ \to 2^+$ transitions), $HF \simeq 2.0$.

The approaches \cite{poe83,Buck:1991,Buck:1992} demonstrated
reliable predictions in our previous experiments where some very
rare $\alpha$ decays were observed at the first time:

-- for $^{180}$W, the calculated values are\footnote{We use here
for calculations the AME2016 $Q_\alpha$ values from \cite{wan17}.}
$T_{1/2}=2.1\times10^{18}$ yr \cite{poe83} and
$T_{1/2}=8.5\times10^{17}$ yr \cite{Buck:1991,Buck:1992} while the
experimental values are in the range of $(1.0-1.8)\times10^{18}$
yr \cite{dan03,coz04,mun14,zde05,bel10};

-- for $^{151}$Eu, the calculated values are
$T_{1/2}=3.6\times10^{18}$ yr \cite{poe83} and
$T_{1/2}=2.9\times10^{17}$ yr \cite{Buck:1991,Buck:1992} with the
experimental half-life near $5\times10^{18}$ yr
\cite{bel07,cas14};

-- for $\alpha$ decay of $^{190}$Pt to the first excited level of
$^{186}$Os ($E_{exc}=137.2$ keV), the calculated values are
$T_{1/2}=4.5\times10^{13}$ yr \cite{poe83} and
$T_{1/2}=2.1\times10^{13}$ yr \cite{Buck:1991,Buck:1992} while the
experimental value is $(2.2\pm0.6)\times10^{14}$ yr
\cite{bel11a}\footnote{We corrected here the original value of
$T_{1/2}=2.6\times10^{14}$ yr \cite{bel11a} calculated for
$^{190}$Pt natural abundance of $\delta=0.014\%$ with the last
IUPAC recommended value of $\delta=0.012\%$ \cite{mei16}.};

-- it is interesting to note that for $^{174}$Hf the calculated
values $T_{1/2}=7.4\times10^{16}$ yr \cite{poe83} and
$T_{1/2}=3.5\times10^{16}$ yr \cite{Buck:1991,Buck:1992} were in
strong contradiction with the old experimental value of
$(2.0\pm0.4)\times10^{15}$ yr \cite{Mac61} (by factor of $17-38$).
However, recent measurements gave new experimental value of
$(7.0\pm1.2)\times10^{16}$ yr \cite{Caracciolo:2020} in a
good agreement with calculations \cite{poe83,Buck:1991,Buck:1992}.

The $T_{1/2}$ values calculated in accordance with
\cite{poe83,Buck:1991,Buck:1992} are presented in
Table~\ref{tab:theory}. In addition, we give the results obtained
here with the semi-empirical formulae of Ref. \cite{den15} which also
were successfully tested with about four hundred experimental
$\alpha$ decays and which take into account non-zero $l$
explicitly. Also, recent calculations of Ref. \cite{Tav20} for Os
isotopes, including transitions to the excited daughter levels,
are presented in Table~\ref{tab:theory}.

\section{Conclusions}

A search for the alpha activity accompanied by the emission of $\gamma$-ray
quanta in naturally occurring osmium isotopes was realized
by an ultra-low background Broad-Energy Germanium
$\gamma$ detector located deep underground at the Gran Sasso
National Laboratory of INFN (Italy). A sample of ultra-pure osmium
with a mass of 118 g, composed of thin osmium slices with an
average thickness of 0.88 mm, was used as a source of the decays.
The isotopic composition of osmium in the sample was precisely
measured with the help of Negative Thermal Ionisation Mass
Spectrometry, that is especially important for the isotope
$^{184}$Os (theoretically the shortest living candidate) whose
representative isotopic abundance was given with a very big
uncertainty of $\pm100\%$ \cite{mei16}.

No $\gamma$-ray quanta expected in the decays searched for were
observed but lower limits on the processes were set at level of
$\lim T_{1/2}\sim 10^{15}-10^{19}$~yr. The half-life limits for
$\alpha$~decays of $^{184}$Os and $^{186}$Os to the first excited
levels of daughter nuclei have been set at 90\% C.L. as
$T_{1/2}\geq 6.8\times10^{15}$~yr and
$T_{1/2}\geq3.3\times10^{17}$~yr, respectively. The limits exceed
substantially the present theoretical estimations of the decays
probabilities that are within $T_{1/2}\sim (0.6-3)\times
10^{15}$~yr for $^{184}$Os and $T_{1/2}\sim (0.3-2)\times
10^{17}$~yr for $^{186}$Os.

A new stage of the experiment is in progress by using an advanced
geometry with the osmium sample placed directly on the BEGe
detector inside its cryostat to increase the detection efficiency
to the low energy $\gamma$-ray quanta expected in the
theoretically fastest decays of $^{184}$Os and $^{186}$Os to the
first excited levels of the daughter nuclei. Obviously, a further
improvement of the experimental sensitivity to the decays with the
highest decay probabilities can be achieved by using
samples of osmium enriched in the $^{184}$Os, $^{186}$Os
and $^{187}$Os isotopes. Observation of other Os isotopes
$\alpha$-instability looks practically problematic taking into
account the very long theoretically predicted half-lives.

\section{Acknowledgments}

D.V.K. and O.G.P. were supported in part by the project
``Investigation of double beta decay, rare alpha and beta decays''
of the program of the National Academy of Sciences of Ukraine
``Laboratory of young scientists'' (the grant number 0120U101838).

\end{document}